\documentclass[12pt]{article}
\usepackage{alltt}

\newcommand{\be}{\begin{equation}}
\newcommand{\ee}{\end{equation}}
\newcommand{\bea}{\begin{eqnarray}}
\newcommand{\eea}{\end{eqnarray}}
\newcommand{\beano}{\begin{eqnarrayno}}
\newcommand{\eeano}{\end{eqnarrayno}}

\def\df{\ifnextchar[{\df@a}{\df@b}}
\def\df@a[#1]#2{{\rm d}^{#1}#2\,}
\def\df@b#1{{\rm d}#1\,}

\def\lsim{\mathrel{\mathpalette\@lsim\sim}}
\def\@lsim#1#2{\lower2pt\vbox{\baselineskip0pt \lineskip.5pt
  \ialign{$\m@th#1\hfil##\hfil$\crcr<\crcr#2\crcr}}}

\def\eqnarrayno{\stepcounter{equation}\let\@currentlabel=\theequation
\global\@eqnswtrue
\global\@eqcnt\z@\tabskip\@centering\def\\{\nonumber\@eqncr}
$$\hbox{}\vcenter\bgroup
  \halign\bgroup\@eqnsel\hskip\@centering
  $\displaystyle\tabskip\z@{##}$&\global\@eqcnt\@ne
  \hskip 2\arraycolsep \hfil${##}$\hfil
  &\global\@eqcnt\tw@ \hskip 2\arraycolsep $\displaystyle\tabskip\z@{##}$\hfil
   \tabskip\@centering&\llap{##}\tabskip\z@\cr}

\def\endeqnarrayno{\nonumber\@@eqncr\egroup\egroup
      \eqno(\theequation)
      $$\global\@ignoretrue}

\newcommand{\method}[1]{{\bf #1}}
\newcommand{\class}[1]{{\tt #1}}
\newcommand{\Cplusplus}{C++}
\newcommand{\parm}[1]{{\it #1}}
\renewcommand{\t}[1]{{\tt #1}} 
\newenvironment{example}{%
  \begin{quote}\begin{alltt}}{%
  \end{alltt}\end{quote}}

\title{A simple C++ library for manipulating scientific data sets as
       structured data }
\author{Christoph Best\\
        {\small Konrad-Zuse-Zentrum f\"ur Informationstechnik}\\
        {\small Takustr.~7, 14189 Berlin, Germany}\thanks{\ \ present address:
        John von Neumann-Institut f\"ur Computing, 52425 J\"ulich, Germany,
        Email: {\it c.best@fz-juelich.de}}}
\date{January 18, 1999}

\begin{document}

\maketitle

\begin{abstract}
Representing scientific data sets efficiently on external storage
usually involves converting them to a byte string representation using
specialized reader/writer routines. The resulting storage files are
frequently difficult to interpret without these specialized routines
as they do not contain information about the logical structure of the
data. Avoiding such problems usually involves heavy-weight data
format libraries or data base systems. We present a simple C++
library that allows to create and access data files that store
structured data. The structure of the data is described by a data type
that can be built from elementary data types (integer and
floating-point numbers, byte strings) and composite data types
(arrays, structures, unions). An abstract data access class presents
the data to the application. Different actual data file structures
can be implemented under this layer. This method is particularly
suited to applications that require complex data structures, e.g.
molecular dynamics simulations. Extensions such as late type binding
and object persistence are discussed.
\end{abstract}

\section{Introduction}

\subsection{Classical input/output mechanisms}

Early programming languages had surprisingly advanced features for
reading and writing data from external memory. For example, COBOL
already had some sort of data definition language,
several file formats and data query statements. It was based on the
notion that the physical representation of the data on external storage
was bit-by-bit identical to the representation in internal memory. As
very early computers had primitive operations to copy data from
external storage into internal memory, this was efficient, but it also
provided a framework for external data representation: records and
fields were usually of fixed length, no separator characters were
necessary, and numbers could be either represented as ASCII or EBCDIC
strings (eventually with an implied decimal point), binary-coded
decimal (storing two digits in a byte) or binary numbers. If you could
read COBOL's data definition language, interpreting file contents was
merely a matter of counting bytes.

In contrast, the data formats used today in scientific computing are
much more flexible, but without human-readable documentation or
careful reading of the source code, it is often impossible to decipher
the contents of a data file. Designing data exchange between different
scientific applications can become a major headache, especially in
small- and medium-sized applications where big input/output libraries
like NCSA's HDF \cite{HDF} or CERN's RD45 \cite{RD45} may be inappropriate.

This problem is in part founded in the design of the C language:
instead of including input/output statements in the language
definition itself, the designers of C decided to implement the whole
input/output functionality in the standard C library, using only
standard function calls. This implies that C cannot provide a
standardized way to store structures - and external data most
frequently is organized in structures, i.e.~records of data
containing dissimilar fields. It is possible to output an arbitrary
structure bit-by-bit, but implementation dependencies such as
alignment rules easily jeopardize compatibility even between different
compiler revisions.

Thus input/output in C programs is usually done manually by writing
explicit code that serializes and reassembles data from a byte stream.
Worse than the additional work and sources of errors associated with
this is the lack of a formal definition of the external data.  The
structure of the data files must either be given in the human-readable
documentation, or---worse---be inferred from the actual source
code. As a consequence, there are no general utilities to manipulate
binary files, e.g.~to produce formatted listings or extract individual 
records and fields.

The situation is only slightly remedied in the more popular scientific
computing language, FORTRAN. While there are input/output statements
in the language, there is no real notion of structured data in FORTRAN
77, so the actual data representation is again encoded in the sequence
of READ and WRITE statements in the code.

\subsection{Object persistence}

In the object-oriented programming paradigma, objects encapsulate data
and the operations that act on the data. A persistent object is an
object with a life-time that extends beyond the life-time of the
program that created it. This means that the object's data must be
stored in external storage, and some operation must be specified to
save and recreate the object from external storage. While this can be
done by implementing the appropriate read and write methods in the
object's class, it is more desirable to have an automated and
standardized way to do so.

A similar problem arises in the design of distributed programming
systems. Architectures like CORBA (the Common Object Request Broker
Architectue \cite{CORBA,Sayegh,MICO}) specify external representations for objects
in terms of their methods, but not of their data. They do so by means
of an Interface Definition Language (IDL) that is abstractly declares
(but not defines) the methods associated with an object class. The IDL
is mapped to appropriate method declarations in specific programming
languages where the implementations of the methods can be provided.
An interoperability protocol defines how methods can be invoked on
data residing on different computers in a network.

A logical extension would be to extend the Interface Definition
Language by a Data Definition Language (DDL). The relation between the
DDL and the actual representation of the data in memory can then be
included in the language mapping, and between the DDL and the actual
representation in external storage in the interoperability
definition.

However, as CORBA is still quite complex and viable implementations
are only recently entering the scientific computing community, and a
standard for a DDL is still missing, one might look around for a poor
man's solution for object persistence. Such a solution should at least 
cover the following requirements:
\begin{itemize}
\item
  The structure of external data should be formally specified, similar
  to the interface definition language of distributed systems.
\item 
  The actual representation of the data on external storage should be
  sufficiently defined by the formal definition of the data structure
  such that the extraction of the data can be performed
  automatically. This corresponds to the interoperability
  specification in distributed programming systems.
\item
  A language mapping or application programmers' interface that makes
  external data easily accessible from application programs.
\end{itemize}
In the following, we present an approach to fulfill these requirements 
using a set of C++ classes.
\begin{itemize}
\item
  The structure of data is formally defined by a type tree using
  elementary and composite data types including structures, arrays,
  and unions (in C parlance). The type tree is built from appropriate
  C++ classes or specified in textual form,
  making use of the appropriate methods for reading and printing type trees.
\item 
  To provide different storage representations, the interface between
  the abstract data layer and concrete representations is specified as 
  an abstract C++ class that can be filled in by different data
  representations. A simple format for arbitrary structured binary
  data is specified.
\item
  From an application program, data objects are accessed using
  a C++ class that represents generic structured data and operations
  performed on such data, such as reading, writing, and extracting
  members of composite data.
\end{itemize}

\section{Implementation}

\subsection{Overview}

The main class provided by the library is \class{SomeData}, an
universal access layer to structured data. To the programmer, each
object of this class represents a structured data item (which may be
elementary or composite). To each object of this class a type tree is
associated that is described by an object of the class
\class{SomeType}.

Before an object can be used, its data type must be specified. This
can be done in three ways:
\begin{enumerate}
\item
  The data type can be specified in the external representation of the 
  data and then be queried by the application program.
\item
  The data type can be explicitly built using the constructors of the
  subclasses of \class{SomeType}:
\begin{example}
    StructType *t1 = new StructType;
    t1->addField("comment",new StringType);
    StructType *tAtom = new StructType;
    tAtom->addField("name",new StringType);
    tAtom->addField("z",new NumType(NumType::i2));
    tAtom->addField("partial_charge", new NumType(NumType::f4));
    ArrayType *tAtoms = new ArrayType(tAtom);
    t1->addField("atoms",tAtoms);
    StructType *tBond = new StructType;
    tBond->addField("from_atom", new NumType(NumType::i2));
    tBond->addField("to_atom", new NumType(NumType::i2));
    tBond->addField("type", new NumType(NumType::i2));
    ArrayType *tBonds = new ArrayType(tBond);
    t1->addField("bonds",tBonds);
\end{example}
\item
  Or the data type can be specified in its textual representation as a 
  string, e.g.
\begin{example}
    const char *typetext =
        "struct \{ "
            "comment : string; "
            "atoms : array of struct \{ "
                "name : string; "
                "z : integer*2; "
                "partial_charge : real*4; "
            "\}; "
            "bonds : array of struct \{ "
                "from_atom : integer*2; "
                "to_atom : integer*2; "
                "type : integer*2; "
            "\}; "
        "\}; ";
      SomeType *t1 = SomeType::parse(typetext);
\end{example}
  and then parsed by a static member function of \class{SomeType}.
\end{enumerate}

After the data type has been specified, a data object must be created.
This is done by a data-set class that manufactures an instance
\class{SomeData}. Data-set classes represent mechanisms where data is
stored and retrieved, e.g.~file formats or databases. The most basic
data-set class is \class{DirectData} that stores the data in a linked
tree in heap memory. While this is of no use by itself, it can be used
by file formats that read their data files in whole and do not wish to
provide their own individual access operators.

One such data-set class is \class{DataFile} that acts as an interface
to text or binary structured files. It provides a method
\method{data()} that returns the data object associated with the file:
\begin{example}
    DataFile DF1;
    DF1.openOut("outfile",t1);
    SomeData D1 = DF1.data();

    DataFile DF2;
    DF2.openIn("outfile");
    SomeData D2 = DF.data();
    SomeType *t2 = D2.typ();
\end{example}
The first group of lines opens a data file for writing using the
previously built type \parm{t1} and acquires the object \parm{D1} to
access the data. The second group opens a data file for input and
obtains a pointer to the data type in \parm{t2}. It may use this
pointer to ascertain that the data has a certain structure.

The main task of \class{SomeData} is to provide data access
methods. An example code fragment manipulating an object \parm{D}
would be:
\begin{example}
    D["comment"] = "blubb blubb";
    SomeData Datoms = D["atoms"];
    for (int i=0; i<10; i++) \{
        SomeData Datom = Datoms[i];
        Datom["z"] = 12;
        Datom["partial_charge"] = 0.0;
        Datam["name"] = "C";
    \}
    SomeData Dbonds = D["bonds"];
    for (int i=0; i<10; i++) \{
        SomeData Dbond = Dbonds[i];
        Dbond["from_atom"] = i;
        Dbond["to_atom"] = (i+1)\%10;
        Dbond["type"] = 1;
    \}
    cout << Datom[0]["z"].getInt() << " " <<
            Datom[0]["name"].getString() << endl;
\end{example}
If \parm{D} is a structured data object, its members can be accessed
by an overloaded indexing operator using either a symbolic name (given 
as a string, for structure data) or an integer index (for array
data). Elementary data types are operated upon either by the assignment
operator, which is properly overloaded for the different data types,
or using access operators like \method{getInt()}.

\subsection{The data type classes}

\subsubsection{Elementary data types}

Three elementary data types are supported
\begin{enumerate}
\item
Signed or unsigned integer numbers with 1, 2, 4, or 8 bytes, e.g.
\begin{example}
integer*4
\end{example}
designates a 4-byte signed integer.
\item
Floating point numbers in IEEE-format with 4, 8, or 16 bytes, e.g.
\begin{example}
real*8
\end{example}
is a 8-byte IEEE 754 floating point number.
\item 
Character strings (that are subject to character-set conversion)
and (opaque) byte strings, e.g.
\begin{example}
string*10
\end{example}
for a 10-byte character string or
\begin{example}
opaque*255
\end{example}
for a 255-byte opaque byte string.
\end{enumerate}

The numerical data types can appear as matrices with arbitrary rank
(i.e.~number of dimensions). Along with the rank, the number of
elements in each dimension must be specified, e.g.
\begin{example}
real*4[100,100]
\end{example}
for a 100$\times$100 floating-point matrix, or
\begin{example}
integer*4[.,2,.]
\end{example}
for a three-dimensional integer matrix whose first and third dimension 
is specified in the data stream.

\subsubsection{Composite data types}

The composite data types are arrays and structures. Arrays a
repetitions of data elements of the same type accessed by integer
indices, while structures are sequences of data elements with
different types accessed either by names or integer indices. A
variation of the structure is the union, in which only exactly one of
many elements of the structure is actually present.

An example of an array is
\begin{example}
  array[100] of integer*4
\end{example}
for an array of 100 4-byte integers. The number of elements may also
be specified in the data stream, e.g.
\begin{example}
  array[.] of array[3] of real*4
\end{example}
is an array in which each element consists of three real numbers. The
number of elements is specified in the data stream.

Structures are specified in the following syntax:
\begin{example}
struct \{
    atoms : array of
        struct \{
            z : integer;
            partial_charge : real;
        \};
    bonds : array of
        struct \{
            from_atom : integer;
            to_atom : integer;
            type : integer;
        \};
    positions : array of integer[3];
    optional velocities : array of integer[3];
\}
\end{example}
This structure has four fields named \t{atoms}, \t{bonds},
\t{positions}, and \t{velocities}. The \t{optional} specifier
indicates that this field may or may not be present in the data stream.

\subsubsection{Implementation}

Any data type is represented by a subclass of class \class{SomeType}.
It provides a method \method{typeP(\parm{t})} that returns \t{true} if
the data type is \parm{t}, where \parm{t} is one of the constants
\t{nilType}, \t{numType}, \t{stringType}, \t{arrayType},
\t{structType}, or \t{unionType} defined in \class{SomeType}, similar
to a dynamic cast. It also defines a virtual method \method{print()}
to print the data type and a static method \method{parse()} to parse a
textual type specification.

The class \t{NumType} defines numerical data. It stores the base type
of data (one of the enumeration constants i1, \t{u1}, \t{i2}, \t{u2},
\t{i4}, \t{u4}, \t{i8}, \t{u8}, \t{f4}, \t{f8}, \t{f16}) and an array
of the dimensions of the matrix. A special value \t{dimFree} is used
for variable-sized dimensions. All this information can be accessed using
accessor methods or specified in the constructor.

Similarly, the class \t{StringType} defines byte-string data. It
stores the number of bytes (or \t{dimFree} for variable-sized data)
and a flag to indicate whether the data is character or opaque. In the 
latter case, it will not be subject to any character-set conversion.

Structured data types are represented by the class \t{StructType}. It
stores an array of fields, each of which is defined by its name, its
type and a flag to indicate whether it is optional. Accessor methods
allow to access fields by index or by name. Structures a constructed
empty, and a method \method{addField(\parm{name},\parm{typ})} is used to
add fields.  Unions are also represented by \t{StructType}, using a
special flag that indicates that the structure is to be treated as a
union.

Arrays use the \t{ArrayType} class which stores the number of elements 
and the type of the elements. Again, the size can be given as the
constant \t{dimFree} to indicate variable-sized arrays.

Parsing of textual type specifications is done by the static member
function \method{parse()} in \class{SomeType}. The syntax is chosen
such that the first word of the type specifications indicates which
subclass the type belongs to. so that the parser can then invoke the 
static member function \method{parse()} in the corresponding
subclass.

\subsection{The data object class}

The class \class{SomeData} provides the basic interface for an application 
to manipulate data. We chose not to duplicate the hierarchy of 
types classes in corresponding data classes but to include accessor
methods for all types of data in a single class. Most accessor
routines for data return objects of the class \class{SomeData}, so
this approach saves the programmer from tedious recasts. When a method 
is called that is improper for the object's data type, it can either
return a null object, or throw an exception.

More important, we wished to include garbage collection by reference
counting in the implementation. This is only possible if the
application program does not use pointers to access objects of the
class \class{SomeData}. Instead, the functionality of the access layer
is split in two parts: Its actual functionality is provided by the
class \class{SomeDataImpl}, that can be subclassed by the different
data representations, while objects of class \class{SomeData} contain
a reference-counted pointer to an object of \class{SomeDataImpl}.
Application programs thus can manipulate objects of class
\class{SomeData} like pointers (or handles). Most methods in
\class{SomeData} either pass through directly to the corresponding
methods in \class{SomeDataImpl} or implement some convenience function 
that can be expressed in terms of these methods. This is especially
advantageous as many operations on \class{SomeData} are expressed by
overloaded functions, e.g.~\method{assign()} to assign any type of
elementary data. Classes that derive from \class{SomeData} and
implement just one of these operations, e.g.~integer assignment, would 
have to implement all overloaded versions of \method{assign()}. In
\class{SomeData}, these operations are seperated into functions like
\method{assignInt()} or \method{assignString()} that provide default
implementations (namely throwing the appropriate exception) and can be 
overloaded individually.

\class{SomeData} provides a method \method{typeP(\parm{t})} to test if 
the data object is of type \parm{t}, and a method \method{typ()} that
returns of pointer to its type (represented by an object of class
\class{SomeType}). It also defines the method \method{copy()} to copy
the contents of one data object into another, assuming that the types
are identical. The assignment operator and copy constructor is defined 
in that way.

For array and structure types, \class{SomeData} contains a convenience
access operator by overloading the indexing operator
\method{operator[]}. If used with a string argument, it accesses a
field of a structure, while with an integer argument, it can be used
on both array and structures to access by index. Its return value is
another object of class \class{SomeData}. This makes accessing
structures and arrays nearly as simple as accessing the corresponding
native data types in \Cplusplus, however, access to fields by name
comes with some performance penalty as the character string must first 
be matched to the names of all fields. This is a general problem with
languages that do not provide a symbol data type (as in LISP): The access
would be faster if the compiler could convert the string into a more
easily manipulated quantity like a 32-bit number that could then be
used to perform a hash or binary search in the field table.

For structures, the indexing operator maps to a method
\method{getField()}. A method \method{nFields()} returns the number of 
fields in the structure, and \method{getFieldName()} the name of each
field. For optional fields, \method{unsetField()} removes the field
while \method{fieldPresent()} checks whether the field is present in
the actual data. For unions, \method{getActiveField()} and
\method{setActiveField()} are used to define which field is used.

For arrays, a method \method{nElements()} returns the number of
elements in the array, and \method{getElem()} is used to access
elements. \method{resize()} can be used to resize the array to a
specified number of elements.

Elementary data are read by the methods \method{getInt()},
\method{getDouble()}, and \method{getString()}. Each returns the
corresponding \Cplusplus data type, or throws an exception if it does
not match the actual data type. A method \method{assign()} with
suitable argument types is used to assign data values.

\subsection{Matrix data}

If the elementary data is a matrix, it is represented in C++ by 
objects of the utility class template \class{Matrix<T>} where \t{T} is 
the elementary C++ data type. Associated with each matrix is 
a shape of class \class{MatrixShape} that stores rank, minimum and
maximum indices in each dimension and information about the storage
layout. The methods \method{getShape()} and \method{setShape()} are
used to manipulate the shape of the data. To access the data itself,
the methods \method{getData()} and \method{assignData()} read and
write the actual representation in memory (as defined by the shape). 

\subsection{Implementation classes}

Actual implementations of data objects are provided by subclassing the 
class \class{SomeDataImpl}. Its member functions are similar to the
member functions of \class{SomeData} without convenience
functions. Its only member field is \method{thetype} which is a
reference-counted pointer to its data type object. 

\subsubsection{Direct representation}

The class \class{DirectData} provides an in-memory representation for
structured data in a linked tree. Its subclasses
\class{DirectStructData}, \class{DirectArrayData},
\class{DirectNumData}, \class{DirectMatData}, and
\class{DirectStringData} are modelled after the subclasses of
\class{StructType} and provide storage for the respective data
types. 

These classes are also used to define a simple text file
representation of the data. Each implements a static member function
\method{read()} to read tokens from a lexical parser and convert them
to an appropriate object. The data format is simple: numbers are
represented naturally, strings are quoted, and arrays and structures
are surrounded by brackets or braces and their elements separated by
(optional) commas. Matrix data are also represented by bracketed lists 
of numbers, with free dimensions specified in front of the
data. Similarly, \class{DirectData} objects know how to print
themselves. 

These methods already allow a complete implementation of the
structured data format. Their main shortcoming is that the data file
must be read as a whole and converted to the \class{DirectData}
format, before it can be accessed by the application.

\subsubsection{Binary-file representation}

Similarly to the textual representation, a sequential binary
stream representation is defined (see sec.~\ref{secDataFormats} for more
details). Data fields follow each other without intervening structure
information, except for length information of variable-sized arrays
and matrices and tag bytes for optional fields and unions.

The class \class{StreamData} provides read access to such
files. Each data item is represented simply by its position in the
file. Elementary data are accessed by reading the bytes at the
specified position into memory. To get a member of a structured data
object, the position of the member is calculated. As the members are
in general of variable size, this usually involves reading all the
members before the requested item (at least so far, that their size can 
be determined). This could be avoided by adding size fields to all
composite data types, but is not implemented in the basoc data
format to keep it as simple as possible.

The big advantage of \class{StreamData} is that only the requested
parts of the data files are held in memory. The binary format for this 
implementation was designed to be as simple as possible and, in
particular, easily writable from FORTRAN programs. However, write
access to such files is not as simple, since this kind of data format
requires the data to be written in sequence. To provide a convenient
representation for writing such data, we once again resort to the
\class{DirectData} implementation and provide a function to write a
\class{DirectData} in binary format to a file. This (as well as
reading data) is performed by a class \class{BinaryDataIO} that
encapsulates the parameters of the binary representation and itself used
C++ streams.

It is, however, desirable to provide a mechanism to write data in
smaller chunks instead of having to store the whole output file in
memory. To do so, we need a way of specifying a part of a data
structure. This is done by the \class{SomeDataIterator} class. An
object of this class is a reference to a \class{SomeData} object
somwhere in a composite data object. The method \method{next()} moves
the pointer to the next object in the tree on the same level. If the
object is composite, \method{hasSubs()} is true and a method
\method{down()} can be used to access the first member object. After
the last object on a level has been retrieved, and end-of-file condition is
raised, and the application can use \method{up()} to go up one level
and continue with \method{next()}. In this way, all objects in a data
tree can be retrieved in exactly the sequence in which they are
written in the data format. 

The method \method{writeBinaryRegion()} in \class{BinaryDataIO} writes 
the data between to \class{SomeDataIterator}s to disk. To fill in the
length information of variable-sized arrays, it keeps a region stack
that contains the byte offsets of all composite data objects that
enclose the current object. 

Using these methods, the class \class{DataFile} that implements data
files provides a method \method{commit(\parm{D})} to commit all data
up to but not including the data item \parm{D} to disk. Before they
are written to disk, the data are stored in a \class{DirectData}
object associated to the data file. After they are written, the
corresponding parts of the data tree are deallocated and the memory
thus freed.

This procedure may be somewhat unsatisfactory as it does provide full
flexibility. However, an implementation that allows filling in data
objects in arbitrary order can be achieved only using a more advanced
data format.

\subsubsection{MallocFile representation}

A more flexible data format can be provided by using an
\class{MallocFile}. This is a flexible block-structured file whose
blocks can be manipulated similarly to the blocks in the C
heap. Blocks can be allocated to arbitrary size and returned to the
free list in any order. Each block is identified by an address and 
be accessed by acquiring a handle object based on the address. As long 
as the handle exists, a copy of the block is locked in memory for
manipulation and written back when the handle is released.

The simplest mapping of structured data to a \class{MallocFile} is to
represent each data object by exactly one block. Composite objects
then contain a list of addresses that identify blocks that represent
their members. When an array is resized, it then suffices to resize
the block that contains the array. Unfilled member fields can be
represented by null pointers and filled at will.

The format can be improved by not allocating a block to each data
objects. Instead, data objects of constant size can be stored directly 
in their parent blocks, in place of the pointers. Thus, an array whose 
objects are of constant size can be stored in a single block. A simple 
rule determines the storage layout: If an object has variable size,
a pointer to the object is stored, otherwise, the object itself shows
up. An object has variable size, if it is a variable-sized elementary
data item (like a matrix with free dimensions), or if it is a
structure with variable-sized or optional fields, or if it is an array 
with a variable-sized element type or with an unspecified number of
elements. 

\section{Data formats}
\label{secDataFormats}

\subsection{Binary data format}

The primary binary data format has been designed with simplicity in
mind. In particular, it can be written from FORTRAN 77 or C simulation 
programs without using the C++ library. The format is basically a
sequential byte string format in which the data fields are written in
the sequence in which they appear when traversing the type tree
without intervening meta-information. The only exception are length
specifications for variable-sized items (matrices or arrays), tag
bytes for optional structure fields, and selector tags for
unions. There are no alignment requirements for data items.

The binary data stream is preceded by an ASCII portion of the file
that contains an identification line with some metainformation,
followed by the textual representation of the data type. Simulation
programs can write this part easily as a string constant. The
following is an example of such a header portion:
\begin{example}
STRUCTURED FILE V0.1 BINARY_BE
#@Date= 18. 3.1998     Time: 15:26
TYPE
struct \{
     molecule_description : struct \{
             molecule_name: string;
             atom_classes : array of struct \{
                              atom_class_id : integer*4;
                              atom_class_number : integer*4;
                              atom_class_name : string;
                                            \};
             atoms : array of struct \{
                       atom_id   : integer*4;
                       atom_name : string;
                                      \};
             bonds : array of struct \{
                       bond_from_id : integer*4;
                       bond_to_id   : integer*4;
                       bond_type    : integer*4;
                                     \};
                                   \};
     timesteps : array of struct \{
                   global_obs : real*4[.];
                   coordinates : real*4[3,.];
                   optional velocity    : real*4[3,.];
                   optional potential : struct \{
                                          bb : real*4[3,2];
                                          data : real*4[.,.,.];
                                               \}
                                 \};
       \};
DATA
\end{example}
The first line starts with the constant "\t{STRUCTURED FILE}" to
identify the file type, followed by a version specification and
optional keyword, here "\t{BINARY\_BE}" indicating that the data are
written in binary format with big-endian byte order. Lines starting
with hash signs are comment lines. The keyword TYPE initiates parsing
of the type tree textual representation. The binary data stream starts 
immediately after the end of the line containing the "\t{DATA}"
keyword.

The following is the specification for the binary data stream:
\begin{enumerate}
\item
Elementary data objects are written in their natural representation
with the byte-order indicated in the identification
line. Floating-point numbers are written in IEEE standard
representation.
\item
Multidimensional data objects are written in FORTRAN order, i.e.~the
first index varies fastest. If any dimension is unspecified in the
data type, the number of elements in this dimension is written as a
4-byte integer in front of the data. This is only done for unspecified 
dimensions.
\item
Character and byte strings are written byte-by-byte. If they are of
unspecified size, the actual size precedes them as a4-byte integer.
\item
Structure and array data are written as consecutive data elements. If
the array size is unspecified in the data type, it precedes the data
as a 4-byte integer.
\item
Optional fields are preceded by a single byte. If this byte is zero,
the field is not present and there follow no data.
\item
Unions are preceded by a 2-byte integer indicating the index (starting 
from zero) of the active field. It is followed by the binary data for
this field only.
\end{enumerate}

\section{Extensions}

\subsection{Named types}
\label{secNamedTypes}

Named types are used to built recursive type trees. In order that a
recursive type tree does not lead to a data tree with infinite
recursion, recursive types usually appear along with unions. An
instructive example of their usage is the following data type that is
used to externalize type trees:
\begin{example}
       typedef TypeDescriptor = union \{
               num : struct \{
                   isFloat : integer*1;
                   size : integer*1;
                   dim : array of integer*4;
               \};
               string : struct \{
                   isOpaque : integer*1;
                   size : integer*4;
               \};
               struct : struct \{
                   isUnion : integer*1;
                   fields : array of struct \{
                       name : string;
                       typ : type TypeDescriptor;
                       isOptional : integer*1;
                   \};
               \};
               array : struct \{
                   size : integer*4;
                   subtype : type TypeDescriptor;
               \};
               named : struct \{
                   name : string;
               \};
           \}
       type TypeDescriptor;
\end{example}
The syntax
\begin{example}
       typedef \parm{typename} = \parm{type}
\end{example}
declares the \parm{typename} to stand in for the \parm{type} wherever
\t{type \parm{typename}} is used. This enables the \t{struct} and
\t{array} variants of the union to reference other type descriptor
trees. The last line of the example is not part of the named-type
definition but declares this type to the application.

\subsection{Late-type binding}

It is not always advantageous to specify the data type separately from
the data. As far as the methods of the \class{SomeData} class are
concerned, it is sufficient for the data type to specify that an
object is a structure or an array, but not what the member fields
are. This decision could be deferred to the moment when the fields are 
actually accessed. The choice of a strong or early type binding that
determines the complete type tree when creating the data object was
motivated by efficiency considerations. In particular, if a data
structure contains arrays of structures, a late-binding data format is 
forced to repeat the field names in each element of the array.

However, in many cases it is desirable to be able to specify that any
data type may appear in a data object, e.g.
\begin{example}
    struct \{
        \dots
        userdata: any;
        lots_of_userdata: array of any;
        \dots
    \};
\end{example}
specifies a structure with a field \t{userdata} that can contain an
arbitrary data type, and an array \t{lots\_of\_userdata} whose elements
can contain any and in particular different data types. This is
especially desirable if the underlying data format is a late-binding
format where the complete type tree can only be retrieved by reading
the whole file.

To be able to store arbitrary data types in a file, the data type must
be specified in the data. One way to do so is to use the textual
representation of the data type and use the \method{parse()} static
member function of \class{SomeType}. Another is to make use of the
data structure from the example in \ref{secNamedTypes}. It contains
all the fields necessary to externalize a complete type tree. The
class \class{SomeType()} provides methods to convert type trees from
and to data objects using this type specification. A data format that
wishes to make use of the \t{any} type will create a data object of
this type, have it filled in by these methods and arrange to have it
written along with the actual data object that contains the \t{any}
data.

To the programmer, the \t{any} type is completely transparent when
reading data. In particular, no instance of \class{SomeData} should
ever be of the \t{any} type when reading. It will only appear in
the type description of members of composite data types. Whenever such 
a member is accessed by the overloaded indexing operator, it will
return the actual data type found in the data file.

When writing data, a data type will be of the \t{any} type until data
is actually written to it. Before you can assign data to an \t{any}
type data object, you must specify the actual data type \parm{t} by
using the \method{actualizeType(\parm{t})} method of \class{SomeData}.
After this method has been called, the data object behaves as if it
was of the type \parm{t}.

Objects of class \t{any} are implemented by means of forwarding. Their
implementation contains a pointer to the target data object whose type
is specified by \method{actualizeType()}. Each method of
\class{SomeDataImpl} forwards to the corresponding method in the
target data object when the forwarding pointer is set. Otherwise, the
\t{any} type object behaves as a null object, illegal to read or
write.

\subsection{Object persistence strategies}

An object persistence mechanism, in theory, relieves the programmer of
the need to program any input/output. Instead, the compiler and/or the 
runtime system take over the task of reading and writing objects from
or to external storage. This could, in theory, be achieved by a
precompiler, but there are several conceptual problems:
\begin{itemize}
\item 
  A class may contain temporary fields, or fields that have
  meaning only when the object is in a certain state. The programmer
  probably does not want these fields to be made persistent,
  especially if these fields are pointer fields.  Handling this
  requires the introduction of a special keyword to mark persistent
  fields in an object.
\item
  C++ relies heavily on pointers which cannot be externalized. They must 
  be replaced by appropriate object handles, and consequently care must
  be taken that all objects that are referenced are also made
  persistent.
\item
  Data structures are often implemented by means of template classes,
  e.g.~the Standard Template Library. These template classes must also 
  be made persistent.
\end{itemize}
Most of these problems stem from the requirement that C++ be
efficient and compatible with C. Languages like Java avoid these
problems by disallowing pointers or template classes.

Is there a way to achieve something similar without using a
precompiler? In Java, a reflection mechanism makes it possible to read 
out the structure of any data type at runtime. Using such a mechanism, 
we could provide a method that takes an arbitrary object and
constructs an appropriate data type for its externalization. Such a
mechanism is currently not available in C++. In particular, there is
no way to get a list of the fields in a structure though this could be 
envisaged as a part of a more general template implementation.

An alternative is the approach taken by Sun's XDR: Define an abstract
data description class whose methods can be used to built data
objects. Then, in each persistent class, add a single method that
takes a pointer to such an abstract data description object and invoke 
the methods corresponding to the fields of the class:
\begin{example}
struct S : public ADRInterface \{
    int a;
    float b;
    struct S2 c;
    struct S3 d[5];
    
    void adr(ADROperation *op) \{
        op->adrInt("a",a); 
        op->adrFloat("b",b);
        op->adrStruct("c",c); 
        op->adrArray("d",d,5);
    \}
\};
\end{example}
This can be further enhanced by using overloaded functions for
describing the data. The methods \method{adrStruct()} and
\method{adrArray()} here expect that their second argument, i.e.~the
data object, is subclassed from \class{ADRInterface} so they can call
the \method{adr()} method to obtain the structure of the objects.

To use such an interface for the \class{SomeData} class, three
different \class{ADROperation}s must be defined: one to obtain the
type tree, and one each for reading and writing. Using these operations,
any object that implements the \method{adr()} method can be
automatically converted into a \class{SomeData} object and back.

\section{Conclusions and outlook}

The library presented here presents a uniform interface for handling
persistent structured data objects in C++. Together with the simple binary
file format defined here, it enables applications to store data on
disk in an easily interpreted but highly flexible format. It relieves
the programmer from the burden to define binary file formats and moves 
the data exchange specification to the level of symbolic field
names. Practical experience shows that this is the most important
feature of the library as it allows to add more fields to a data
format without interfering with already existing programs.

For reasons of performance, more sophisticated data representation can
be added to the library. One such, based on the MallocFile, has been
discussed above, others could include interfaces to established data
formats like HDF or to relational or object-oriented data bases using
SQL.

As everyone talks about object-oriented programming, extending the
library in this direction can be achieved by adding method members to
structures, making it possible to invoke operations on
\class{SomeData} objects. Return type and argument lists of such
methods can again be specified by abstract type trees and passes as
\class{SomeData} objects. The actual code executed by a method can be
hidden away from the application by the data-set classes, thus making
it possible to invoke e.g.~a Java method from a C++ program. However,
this is exactly the feature provided by distributed object systems
like CORBA or ILU \cite{ILU}.

This also shows that abstract type trees and a uniform interface to
structured data is not restricted to object persistence. Java recently
introduced a reflection interface \cite{Java2} that makes type trees
of Java objects accessible from Java code at run-time, and as not
everybody can move to Java, especially in scientific computing, a
simple kitchen-sink solution like the one presented here might ease
some of the everyday problems in C++ programming.

{\bf Acknowledgements:} I would like to thank Frank Cordes for his
effort at integrating the data format in a molecular dynamics
application, and Daniel Runge, Johannes Schmidt-Ehrenberg, and
Hans-Christian Hege for discussion.

\end{document}